\pdfoutput=1
\documentclass[sigconf]{acmart}

\usepackage{booktabs} 
\acmConference[ICVGIP’25]{Indian Conference on Computer Vision, Graphics and Image Processing, Mandi, India}{December 2025}{Mandi, India}
\begin{document}
\title{Secure Audio Embedding in Images using Nature-Inspired Optimization}

\author{Aman Kumar, Ankit Chaudhary}
\affiliation{
  \institution{Jawaharlal Nehru University}
  \department{Cybernetics Research Group, School of Engineering}
  \city{New Delhi}
  \country{India} \break
\href{mailto:amanku33_soe@jnu.ac.in}{amanku33\_soe@jnu.ac.in},
\href{mailto:ankitchaudhary@jnu.ac.in}{ankitchaudhary@jnu.ac.in}}

\begin{abstract}
In today’s digital world, protecting sensitive data is very essential. Steganography hides the existence of secret data instead of its content, providing better security for multimedia communication. This paper proposes a new technique for hiding audio files inside images using the Least Significant Bit (LSB) method optimized by the Harris Hawks Optimization (HHO) algorithm. HHO is a nature-inspired metaheuristic that imitates the hunting behavior of Harris hawks to find optimal pixel positions for embedding data. The proposed method is evaluated using Peak Signal-to-Noise Ratio (PSNR), Structural Similarity Index (SSIM), and Mean Square Error (MSE). Experimental results show that HHO achieves better image quality, robustness, and embedding capacity compared to existing methods.
\end{abstract}


\maketitle

\section{Introduction}
The security of digital information has become a key challenge in modern communication systems. Cryptography protects message content, while steganography conceals the communication itself \cite{saxena2021digital}. Image steganography embeds secret data into image pixels in a way that changes are invisible to the human eye\cite{Gupta2018MetaHeuristic, Nokhwal2023MetaEmb, Nokhwal2024SecureInfoEmbedding}. The Least Significant Bit (LSB) technique is one of the most used methods for data embedding \cite{nandhini2016mlsb}, \cite{karim2011new}.

However, traditional LSB substitution is vulnerable to statistical steganalysis and can expose hidden data \cite{ashu2014glcm}. To overcome this issue, researchers have applied optimization algorithms such as Genetic Algorithm (GA), Particle Swarm Optimization (PSO), and Ant Colony Optimization (ACO) to improve embedding efficiency and minimize distortion \cite{wanga2001image, kennedy1995particle, dorigo2006ant}.

In this study, we propose a Harris Hawks Optimization (HHO)-based method to enhance audio hiding in images. HHO provides faster convergence and higher-quality stego images than other metaheuristics \cite{heidari2019harris}.

\section{Methodology}
\begin{figure}[h]
\centering
\includegraphics[width=0.85\linewidth]{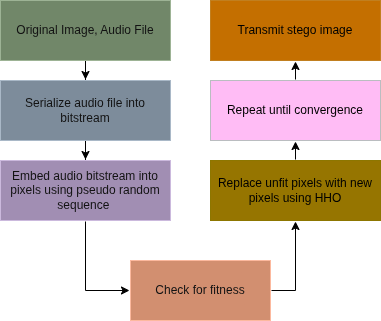}
\caption{\label{fig:3}Block diagram of Proposed Method}
\end{figure}

A block diagram illustrating the workflow of the proposed HHO-based steganography process including audio preprocessing, HHO-based embedding, and stego image generation is shown in Figure \ref{fig:3}.

\subsection{Audio and Image Preprocessing}
We convert the input WAV audio into a bitstream. We use standard test images. Each RGB channel is considered for embedding, but candidate pixels are limited to high-variance (edge or textured) blocks to reduce visual distortion \cite{jawahir2015audio}.

\subsection{HHO for Pixel Selection}
\begin{itemize}
    \item \textbf{Solution encoding:} Each hawk (candidate) encodes a list of pixel indices (or coordinate pairs) where audio bits will be embedded.
    \item \textbf{Fitness evaluation:} For each candidate embedding pattern, we simulate embedding with LSB and compute PSNR, SSIM, and MSE. The combined fitness is given by eq. (1):
    \begin{equation}
    \label{eq12}
    Z(x,y) = \alpha*SSIM + (1-\alpha)*PSNR/100 
    \end{equation}

    Where $\alpha$  is weight constant, in our case, it is 0.5 as we are considering 50\% SSIM and 50\% PSNR.

    \item Phases in HHO (per Heidari et al. \cite{heidari2019harris}):
    \begin{enumerate}
        \item \textbf{Exploration:} Hawks randomly probe solutions to explore the embedding space.
        \item \textbf{Transition:} Hawks adjust strategies as prey (optimal) location is approximated.
        \item \textbf{Exploitation / Besiege:} Local refinement around promising embedding sets (soft besiege, hard besiege).
    \end{enumerate}
    \item \textbf{Update rules:} Equations for position updates include volatile jumps or encirclements depending on energy parameter $E$.
    \item \textbf{Termination and selection:} The algorithm runs until a maximum iteration (e.g. 200) or no significant improvement, then the best hawk’s solution is chosen.
    
\end{itemize}

\subsection{LSB Embedding Scheme}
After selecting pixel locations, we perform adaptive LSB substitution: the least significant bit(s) of the selected pixels in RGB channels are replaced by bits from the audio bitstream.If more audio bits than available positions, we either (a) reuse patterns, or (b) embed across multiple color channels— whichever yields acceptable distortion.

\section{Results}

The HHO-based method was tested on standard benchmark images. The embedded audio signal was stored in WAV format and hidden in RGB channels using LSB substitution.

\begin{figure}[h!]
\centering
\includegraphics[width=0.90\linewidth]{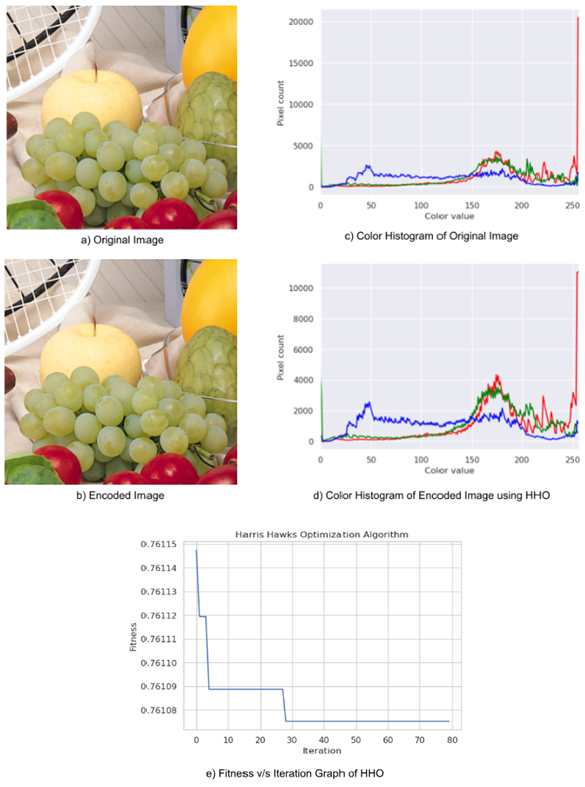}
\caption{\label{fig:8}Harris Hawks optimization  steg results  are shown in images a,b,c,d,e above}
\end{figure}

Performance metrics were calculated as follows:
\begin{itemize}
    \item MSE \cite{thu2008scope}: minimized significantly compared to other algorithms.
    \item PSNR \cite{hore2010image}: above 55 dB for all tested images.
    \item SSIM \cite{wang2004image}: approximately 0.999 for high-resolution images.
\end{itemize}
.

In Figure \ref{fig:8} histogram analysis indicated negligible differences between cover and stego images, confirming that visual quality was preserved. The RGB pixel distribution remained stable, as only the least significant bits were modified.

Table \ref{tab:Iter} summarizes the comparative results of all algorithms evaluated in the original study. HHO required only 200 iterations to achieve convergence, outperforming other algorithms such as FA (1000 iterations) and CS (800 iterations).

\begin{table}[h]
\centering
\begin{tabular}{|c|c|c|c|}
\hline
Sr. No. & Name & Time (150 It.)(in min) & Iterations to optimal \\\hline
1 & FO & 250 & 500 \\
2 & HHO & 265 & 200 \\
3 & CS & 270 & 800 \\
4 & FA & 70 & 1000 \\
5 & ALO & 255 & 700 \\\hline
\end{tabular}
\caption{\label{tab:Iter}Time taken and avg number of iterations taken by different algorithms}
\end{table}

HHO achieved optimal embedding within 200 iterations. Histogram analysis also confirmed minimal visual difference between cover and stego images

The experimental results verify that the proposed HHO-based embedding technique ensures imperceptibility, robustness, and security—key aspects of effective steganographic communication.

\section{Conclusion}
The proposed HHO-based technique intelligently selects optimal pixel locations for embedding audio bits, mimicking the cooperative hunting behavior of Harris hawks to explore and exploit the solution space efficiently. This optimization significantly reduces image distortion while maximizing data-hiding capacity. Compared to traditional LSB methods and other optimization algorithms such as Particle Swarm Optimization (PSO), Firefly Algorithm (FA), and Cuckoo Search (CS), the HHO-based approach demonstrated faster convergence and better embedding performance.

Experimental results confirmed that the proposed model provides superior visual quality and security. The Peak Signal-to-Noise Ratio (PSNR) remained above 55 dB, indicating minimal degradation of image quality. Similarly, the Structural Similarity Index (SSIM) was nearly 0.999, suggesting that the stego image was almost identical to the original cover image. Additionally, histogram analysis verified that the pixel intensity distributions of cover and stego images were nearly identical, confirming the imperceptibility of the embedded data.

Another major advantage of this approach is its computational efficiency. HHO achieved optimal embedding performance in fewer iterations compared to other nature-inspired algorithms. This efficiency makes the method practical for real-world applications where both speed and security are essential.

\bibliography{ICVGIP-Latex-Template}
\bibliographystyle{unsrt}

\end{document}